\documentclass[floatfix,superscriptaddress,twocolumn,showpacs,preprintnumbers,amsmath,amssymb,pre]{revtex4}
\usepackage{graphicx}
\def\Tr{{\text{Tr}}}
\def\H{{\text{H}}}
\def\ima{\imath}
\def\>{\rangle} 
\def\<{\langle} 
\newcommand{\ie}{\textit{i.e.} }
\newcommand{\fref}[1]{Fig.~\ref{#1}}
\newcommand{\eref}[1]{Eq.~(\ref{#1})}
\newcommand{\mcH}{\mathcal H }

\graphicspath{{eps/}}
\begin{document}
\title{Non-universal level statistics in a chaotic quantum spin chain}
\author{Carlos Pineda}
\email{carlospgmat03@gmail.com}
\affiliation{Instituto de Ciencias F\'{\i}sicas, Universidad Nacional Aut\'onoma de M\'exico, M\'exico}
\affiliation{Instituto de F\'{\i}sica, Universidad Nacional Aut\'onoma de M\'exico, M\'exico}
\affiliation{Centro Internacional de Ciencias, Cuernavaca, M\'exico}
\author{Toma\v z Prosen}
\email{tomaz.prosen@fmf.uni-lj.si}
\affiliation{Physics Department, Faculty of Mathematics and Physics, University of Ljubljana, Ljubljana, Slovenia}
\date{\today}
\begin{abstract}
We study the level statistics of an interacting multi-qubit system, namely the
kicked Ising spin chain, in the regime of quantum chaos. Long range
quasi-energy level statistics show effects analogous to the ones observed in
semi-classical systems due to the presence of classical periodic orbits, while
short range level statistics display perfect statistical agreement with random
matrix theory.  Even though our system possesses no classical limit, our result
suggest existence of an important non-universal system specific behavior at
short time scale, which clearly goes beyond finite size effects in random
matrix theory.
\end{abstract} \pacs{05.30.-d,05.45.Mt,05.45.Pq}
\keywords{many body system, quantum dynamics, quantum chaos, classical chaos, quantum statistics,periodic orbits}
\maketitle
\section{Introduction}

One of the key discoveries of quantum chaos has been the so-called quantum
chaos conjecture, originally proposed in Refs.~\cite{qcc}. It claims that
even simple non-integrable quantum systems, whose  
dynamics is sufficiently
complex (say, dynamically mixing in the classical limit), possess quantum
fluctuations which can be described by a universal ensemble of random matrices
without any free parameters \cite{mehta}. Although a strict mathematical proof of
this conjecture is still missing, its theoretical understanding has recently
been considerably deepened \cite{qcctwo}. Still, it is known from the early
years of quantum chaos \cite{berry85}, that level fluctuations exhibit
universal features only on sufficiently small energy scales, or long time
scales, whereas one obtains system specific non-universal features on long
energy scales (short time-scales) which can be usually understood and computed
in terms of classical orbits.

Therefore it seems that the picture is quite complete and satisfactory for
systems possessing a well defined classical limit. But what about simple
systems which do not have a classical limit, e.g., systems of interacting
fermions, or systems of interacting qubits?  In such systems, dynamical
complexity can be reached in the thermodynamic limit of many interacting
particles \cite{prosenpre99b}.  In some exactly solvable cases formal similarities between the
thermodynamic limit and the semi-classical limit can be established
\cite{prosenpre99}.  For example, one may start by considering simple, {\em
non-integrable}, many-particle Hamiltonians with local interaction which are
specified by only a few ({\em non-random}) parameters.  Can quantum spectral
fluctuations of such systems be described by universal ensembles of random
matrices? If yes, what are the energy scales of such universality?  Is there a
breaking of universality at sufficiently large energy ranges? How does
the universality breaking scale in the thermodynamic limit?  In this paper we address
these questions in a simple dynamical system, namely an
Ising chain of spin $1/2$ particles on a 1d ring, kicked periodically with a
homogeneous, tilted magnetic field.  We performed careful numerical
calculations of quasi-energy spectra and their statistical analysis.  For
appropriate values of model's parameters, corresponding to strong integrability
breaking, we indeed find both, the {\em universality regime} for sufficiently
small energy scales, where no statistically significant deviations from random
matrix prediction of infinitely dimensional Circular Orthogonal Ensemble (COE)
could be detected, 
and a {\em non-universality regime} for large energy scales (or small times,
corresponding to few kicks), where clear, statistically significant deviations
from Random Matrix Theory (RMT) prediction have been found. Most notably, our
analysis shows that the spectral form factor exhibits significant deviations
from RMT at the time scale corresponding to one or few kicks (Floquet periods).
This result could be intuitively understood as a qubit analogy of  ``shortest
periodic orbit'' correction, but its precise theoretical understanding is at
present open.

\section{The system}

The system we study is a {\em kicked Ising chain} (KIC) \cite{prosenKI}, namely
a ring of $L$ spin $1/2$ particles which interact with their nearest neighbors
via a homogeneous Ising interaction of dimensionless strength $J$ and being
periodically kicked with a homogeneous magnetic field of dimensionless strength
$\vec{b}$. During the free evolution, \ie between the kicks, the system evolves
with the unitary propagator
\begin{equation} \label{eq:Ising}
	U_\text{Ising}(J) =\exp \left( -\ima J \sum_{j=0}^{L-1} \sigma ^z_j \sigma^z_{j+1} \right),
\end{equation}
and the action of the kick is described by the unitary operator
\begin{equation} \label{eq:kick}
	U_\text{kick}(\vec{b})=\exp\left(-\ima  \sum_{j=0}^{L-1} \vec{b} \cdot\vec{\sigma}_j \right),
\end{equation}
with $\sigma_j^{x,y,z}$ being the Pauli matrices of particle $j$ and
$\vec{\sigma}_j=(\sigma_j^x,\sigma_j^y,\sigma_j^z)$. The Floquet operator for
one period is thus 
\begin{equation} \label{eq:floquet}
	U_\text{KI}=U_\text{Ising}(J)U_\text{kick}(\vec{b}).
\end{equation}
We must also impose periodic conditions in order to close the ring:
$\vec{\sigma}_{L} \equiv \vec{\sigma}_0$.  During this paper we shall use the
so called computational basis, which is composed of joint eigenstates of
$\sigma^z_j$.  This set of basis states can be written as $ S=\left\{ | m_0 m_1
\dots m_{L-1}\>,\, \textrm{with}\, m_j \in \{0,1\}\right\}$.

In order to understand the spectrum one must first discuss the symmetries in
the system.  Let us start with the translational symmetry. 
The corresponding operator $T$ is defined on the computational basis as
$ T | m_0 m_1 \cdots m_{L-1}\> = | m_{L-1} m_0 \cdots m_{L-2}\> $, and is
extended to the entire
Hilbert space by linearity.  The action of this operator is to rotate the
particles in the ring by one site.  The eigenvalues of $T$ are $\exp(2\pi\imath
k/L)$ with $k\in \mathbb{Z}_{/L}$. Hence the Hilbert space is foliated into $L$
subspaces $\mathcal H=\bigoplus_{k\in \mathbb{Z}_{/L}}\mathcal H_k$.
The evaluation of the dimensionality of each of these subspaces is described
in Appendix~\ref{sec:dimhilbert}.  The evolution operator
(\ref{eq:floquet}) is translationally invariant, and hence $[U_\text{KI},T]=0$. 

The next symmetry is an external reflection $R$.  Its action on
the basis $S$ reads $ R | m_0 m_1 \cdots m_{L-1}\> = | m_{L-1} m_{L-2} \cdots
m_{0}\>$, and also $[U_\text{KI},R]=0$.  The two symmetries, $T$ and
$R$, do not commute. It must be noticed that if $|\psi\> \in \mathcal H_k$,
then $R|\psi\> \in \mathcal H_{-k}$. This provides an additional symmetry
within the subspace $\mathcal H_{0}$ (and $\mathcal H_{L/2}$ for even
$L$). Thus these marginal subspaces are regarded as ``special'' and
have slightly different properties than the rest. We shall not consider them
for the purpose of statistical analysis in this article.   

Finally, we define the anti-unitary symmetry $\mathcal K'$. It acts as a mirror
reflection within each spin with respect to the plane that contains both $\vec
b$ and the unitary vector in the $z$ direction (the direction of the Ising
interaction). We can rotate our coordinate system around the $z$ axis in each
qubit so that $\vec b$ only has components in the $x$ and $z$ directions. Then,
$\mathcal K'$ is simply complex conjugation, provided that $\sigma_x$ and
$\sigma_z$ are set real, as is the usual choice.  This symmetry operation also
changes the sign of the momentum. Composing $\mathcal K=\mathcal K' R$, we
arrive to an anti-unitary symmetry that preserves the momentum \ie an
anti-unitary symmetry within each $\mathcal H_k$.  

Concluding, $U_\text{KI}$ has a rotational symmetry $R$ that foliates the
spectrum into $L$ different sectors; the sector $k$ has identical spectrum as the sector
$-k$.  Hence for a fixed number of qubits we expect to have a maximum of
$(L-1)/2$ relevant sectors, each sector having a Hilbert space dimension ${\cal
N}\approx 2^L/L$. Since each sector has an internal anti-unitary symmetry, we
shall compare the statistical properties of system's spectrum to those of
Dyson's Circular Orthogonal Ensemble \cite{mehta} of random matrices, of
appropriate dimension.

\begin{figure}
	\begin{center} \includegraphics{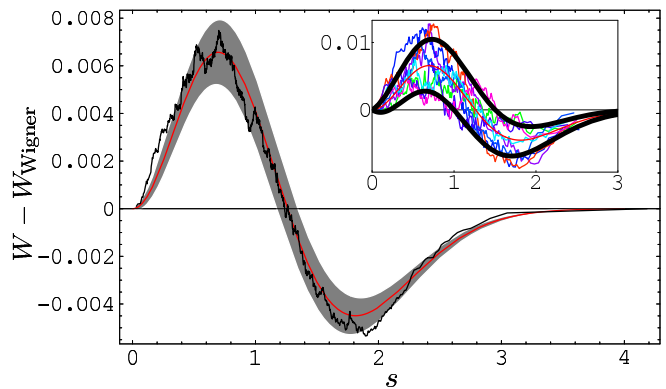} \end{center}
	\caption{(Color online) We study the behavior of the integrated nearest
	neighbor spacing distribution, $W(s)$. The noisy curve shows the
	difference between the numerical data for 18 qubits, averaged over the
	different relevant $\mcH_k$ spaces, and the Wigner surmise.  The smooth
	(red) curve is the difference between infinitely dimensional COE solution and the
	Wigner surmise. The expected standard deviation due to the finite size
	of the spectrum $\sigma_W$ [see \eref{eq:sigmaW}] is also indicated as
	the shaded area surrounding the RMT result. In the inset we present a
	similar figure with the results for each of the $\mcH_k$ subspaces
	plotted separately, together with the error associated with each
	individual spectrum, as the thick black curves.  (Different colors
	represent different sectors, according to the coding shown in
	\fref{fig:sigma2normal}.) }
	\label{fig:Ws}
\end{figure}

For the rest of the presentation we fix parameter values of our system
$J=0.7$, $\vec{b}=(0.9,0,0.9)$ for which the integrability of the model is
strongly broken. We believe that for these parameter values the system is a
generic representative of quantum chaos.  Using highly optimized numerical
methods (see appendix \ref{basis}) we have been able to diagonalize
the model accurately for sizes up to $L=18$ qubits. The eigenvalues of the
Floquet propagator $U_{\rm KI}$ have been written as $\exp(-\ima\varphi_n)$,
where $\varphi_n$ are known as quasi-energies, and have been grouped with
respect to the known quasi-momentum $k$.  Statistical analyses of desymmetrized
quasi-energy spectra $\{\varphi_n\}$ and their interpretation are given in the
following sections. In order to compare with the RMT formulae we normalize the
quasi-energies, i.e. write $s_n=\frac{{\cal N}}{2\pi}\varphi_n$, in order to
have mean level spacing equal to one (${\cal N}$ denotes the dimension of the
Hilbert space).  Similarly, $s=\frac{{\cal N}}{2\pi}\varphi$ will denote the
spectral variable, unfolded to a unit mean level spacing.

\section{Universality regime}

\begin{figure}
	\begin{center} \includegraphics{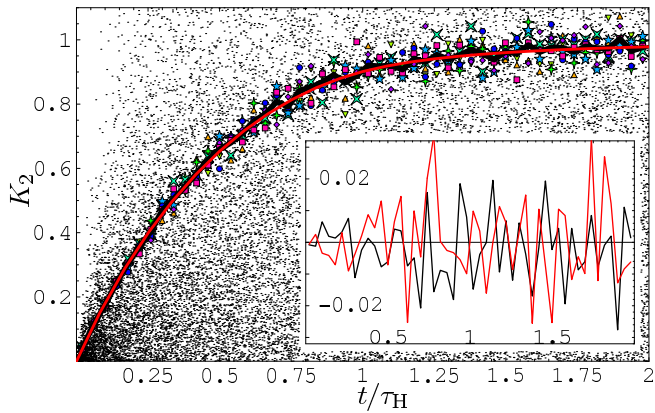} \end{center}
	\caption{(Color online) In this plot we show the behavior of the form
	factor for $18$ qubits. The black dots show its value at integer times
	for the quasi-momentum sector $k=1$. In order to appreciate clearly its
	behavior it is necessary to perform a windowing over short ranges of
	time ($\tau_\H/25$). The results for each of the $k$-spaces are shown
	according to the symbol scheme in \fref{fig:sigma2normal}. The average
	over the different spaces as well as the theoretical curve is plotted
	as a black and red line, respectively. In order to compare with the
	ensemble fluctuations, we plot in the inset the difference from
	the theoretical prediction of both, the spectra for the KIC (in black),
	and the spectra of an equal number of random realizations of COE
	members with the same dimension (in red).}
	\label{fig:K2short}
\end{figure}

Let us first analyze the most commonly studied spectral statistics of chaotic
systems, that is the nearest neighbor level spacing distribution $P(s)$.
$P(s){\rm d}s$ is the probability that the distance between two nearby
(unfolded) quasi-energies is between $s$ and $s+{\rm d}s$.  $P(s)$ has been
computed for the KIC and compared to the exact random matrix COE result
(computed from Pade approximants \cite{dietz1990}) with satisfactory results
(not shown). However, since the details of such plot depend on the size of the
binning of histograms, we prefer to show the cumulative (integrated) level
spacing distribution $W(s) = \int_0^s {\rm d} s' P(s')$.  In fig.~\ref{fig:Ws} we
show a comparison of $W(s)$, both for the KIC and the exact infinitely-dimensional 
COE result, with the Wigner surmise $W_{\rm Wigner}(s)=1-\exp(-\pi s^2/4)$.  
The expected statistical fluctuation of cumulative
probability can be estimated \cite{prosenrobnik1993} as 
\begin{equation} 
	\sigma_W = \sqrt{\frac{W(1-W)}{\cal N}}
	\label{eq:sigmaW}
\end{equation} 
and gives a very realistic estimate of actual fluctuations
of our dynamical system.  We plot the results both for individual
quasi-momentum $k$ subspaces, and averaged over all $k$. In conclusion, based
on the nearest neighbor level spacing distribution we find no significant
deviations from universality, i.e. from COE model statistics.

\begin{figure}
	\begin{center} \includegraphics{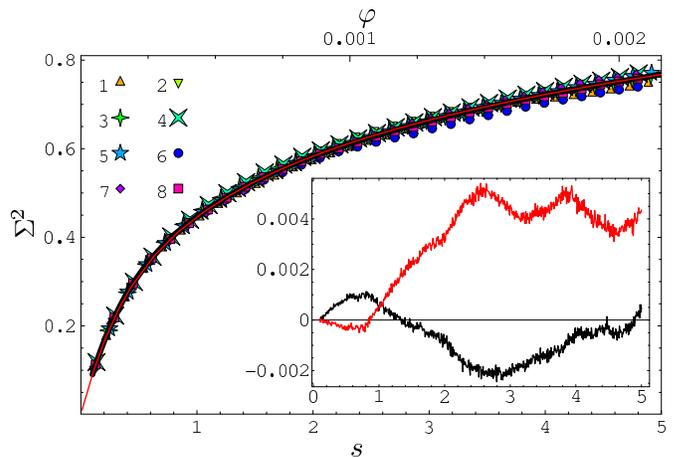} \end{center}
	\caption{(Color online) We observe the variance $\Sigma^2$ for each
	symmetry sector (as different symbols)  and its average (thick black
	curve), for 18 qubits.  Note that lower abscissa indicates the unfolded
	spectral variable, while the upper abscissa the one without unfolding.
	The average curve is almost indistinguishable from the theoretical
	value (thin red curve).  In the inset we compare the deviation of the
	averaged $\Sigma^2$ for both the KIC (in black) and the COE (in red)
	from the theoretical value $\Sigma^2_\textrm{COE}$. No qualitative
	difference is observed. }
	\label{fig:sigma2normal}
\end{figure}

Further on, we have studied other statistical measures of quasi-energy spectra,
which are more sensitive to long-range spectral correlations, namely the number
variance and the form factor \cite{mehta}.  The spectral form factor $K_2$, is
defined for discrete time $t$ as $K_2(t/\tau_{\rm H}) = |\Tr\ U^t|^2/{\cal N}$,
and for infinitely dimensional COE has the form
\begin{equation}
	K_{2,\rm COE}(\tau)=\begin{cases} 
		2|\tau|-|\tau| \ln(2|\tau|+1) & {\rm if }\; |\tau|<1 \\ 
		2-|\tau|\ln \frac{2|\tau|+1}{2|\tau|-1} 
			&{\rm if }\; |\tau|\ge 1 \end{cases}.
	\label{eq:k2coe}
\end{equation}
$\tau_{\rm H} = {\cal N}$ denotes the discrete Heisenberg time, namely the
number of kicks in which the average quasi-energy level separation grows to
$2\pi$.  Asymptotic finite dimension corrections to the form factor have been computed, and
for small $\tau \ll 1$, the result reads
\begin{equation}
	K_{2,\rm COE}(\tau,{\mathcal N})
	=\left[1+\frac{1}{\mathcal N}+
	  \mathcal{O}\left(\mathcal N^{-2}\right) \right]K_{2,\rm COE}(\tau).
	\label{eq:k2coefinite}
\end{equation}
The number variance $\Sigma^2(s)$ gives the variance of the number of levels in
an unfolded spectral interval of length $s$.  The RMT formula for infinitely
dimensional COE predicts a monotonically increasing variance $\Sigma^2_{\rm
COE}(s)=(2/\pi^2)[\ln(2\pi s) + 1 + \gamma - \pi^2/8] + {\cal O}(s^{-1})$,
where $\gamma = 0.5772\ldots$ is the Euler constant \cite{mehta}.  However, for
a finite spectrum of ${\cal N}$ quasi-energy levels this is not possible, since
when the energy difference reaches the range of the spectrum the
number of levels counted will always be the maximum and hence the number
variance will be zero.  For arbitrary finite dimension ${\cal N}$, there is an
exact relationship \cite{dodo} between $\Sigma^2$ and $K_2$ that accounts for
the finite range of the spectrum:
\begin{equation} \Sigma^2(s,{\cal N})=
	\frac{2\mathcal N}{\pi^2}\sum_{m=1}^\infty \frac{1}{m^2} 
	\sin^2 \left( \frac{m\pi s}{\cal N} \right)
		K_2\left(\frac{m}{\tau_\H}\right). 
	\label{eq:finiteS2}
\end{equation}
Truncating the above series at finite  
$m$ with the form factor
given by eq.  (\ref{eq:k2coefinite}) provides an excellent asymptotic
approximation to the COE number variance for finite ${\cal N}$.

\begin{figure}
	\begin{center} \includegraphics{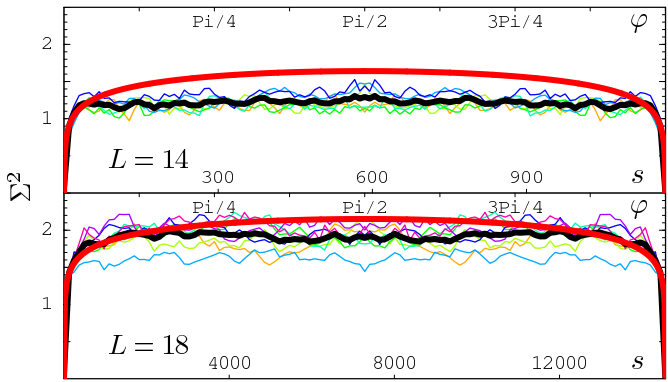} \end{center}
	\caption{We plot $\Sigma^2$ for the KIC for $L=14$ (upper plot) and
	$L=18$ (lower plot). Different sectors are plotted
	using thin colored curves (with the same color coding as in
	\fref{fig:sigma2normal}) and the average value as the thick black curve. 
	The theoretical (COE) prediction for ${\cal N}$ dimensional
	circular random matrices [eq. (\ref{eq:finiteS2})] is also plotted as a smooth
	red curve.  We observe saturation of the stiffness for the KIC,
	characteristic of semi-classical systems.}
	\label{fig:s2Big}
\end{figure}

In figure \ref{fig:K2short} we compare the spectral form factor of the KIC with
the infinitely dimensional COE,  on a global time scale (on the order of
Heisenberg time $\tau_{\rm H} = {\cal N}$).  Of course, since the form factor is
not self-averaging we had to perform some averaging over short time-windows in
order to wash away the statistical fluctuations.  We find no notable deviation
from the COE.  In order to estimate the expected fluctuations due to a finite
sample of systems (namely a set of $\approx L/2$ quasi-momenta $k$) we have also
generated a similar average over the same number of random matrices of
equivalent size.  In the inset we plot the deviations of the form
factor computed for the KIC, and the corresponding finite average over random
members of the COE, from the exact RMT prediction.  We observe that both behave
similarly.  In addition, we find very good agreement for the number variance
$\Sigma^2$ of the KIC with the infinitely dimensional COE on short and
intermediate spectral ranges $s < 10$ (see fig.~\ref{fig:sigma2normal}). In the
inset the finite size fluctuations are also compared with the ones computed
from appropriate finite samples of finite dimensional COE.  Again we observe
agreement.

\section{Deviations from universality}

\begin{figure}
	\begin{center} \includegraphics{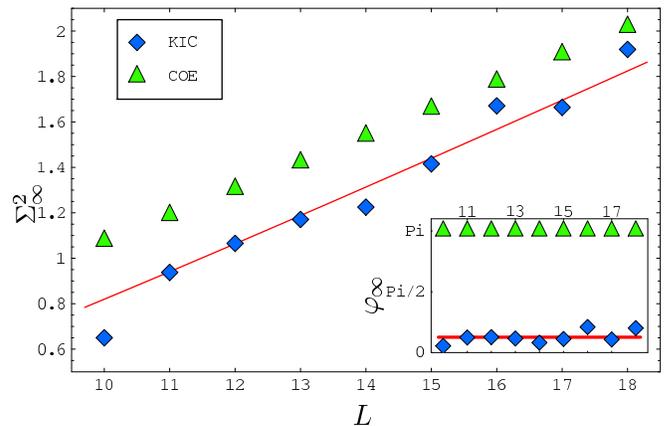} \end{center}
	\caption{(Color online) We see the dependence of $\Sigma^2_\infty$ as a
	function of the number of qubits $L$, both for the KIC and the COE of
	appropriate dimension.  In the inset we plot $\varphi_\infty$.  The
	constant line $\varphi_\textrm{c}=0.39$ is shown in red, as well the
	plot of  $\Sigma^2_{\rm COE}({\mathcal N} \varphi_\textrm{c}/2\pi,{\mathcal N})$ in
	the main panel.}
	\label{fig:plato}
\end{figure}

As explained in the previous section, we expect that the number variance for a
finite spectrum reaches a maximal value at $\varphi=\pi$ (i.e. $s \sim {\cal
N}/2$). Actually as already mentioned, one can compute a good analytical
approximation to COE averages of number variance for finite ${\cal N}$ using
eq.~(\ref{eq:finiteS2}), and the saturation can be understood as a consequence
of discreteness of time in the sine-like transformation on the RHS of
(\ref{eq:finiteS2}).

Do the spectra of KIC in the regime of quantum chaos follow the same
saturation as would be expected for typical members of the COE or not?  We have
performed detailed numerical checks of these questions and report the results
in the following figures.  In fig.~\ref{fig:s2Big} we plot the number variance
for two different number of qubits (14 and 18), for the KIC.  We find a very
clear and notable difference: the data for the KIC tends to saturate at
different, lower value of the unfolded spectral parameter $s_{\infty} \ll {\cal
N}$, than COE, which typically saturate only at $s \sim {\cal N}/2$.
Furthermore, the plateau is quite notorious. 
In the next plot (fig.~\ref{fig:plato}) we have determined the
saturation threshold $s_\infty$ and the saturation value $\Sigma^2_\infty =
\Sigma^2(s_\infty)$ as a function of the number of qubits $L$.

Numerical results suggest that $s_\infty \approx 0.062 {\cal N} \propto
2^L/L$, namely that $s_\infty$ is proportional to ${\cal N}$ though it is
smaller by a large constant factor. The saturation value $\Sigma^2_\infty$ thus
increases logarithmically with ${\cal N}$, or linearly with $L$.

\begin{figure}
	\begin{center} \includegraphics{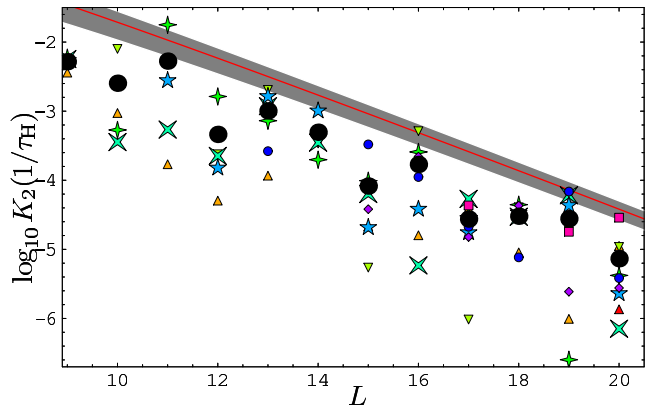} \end{center}
	\caption{We plot the value of $K_2$ evaluated at the first kick, for
	different quasi-momentum sectors (symbols according to
	fig.~\ref{fig:sigma2normal}) and averaged over all quasi-momenta
	(filled circles).  The red line indicates the theoretical COE value
	surrounded by one expected standard fluctuation (according to
	theoretical COE fluctuation) indicated by gray area. The average is
	systematically bellow the expected RMT value.}
	\label{fig:K2at1}
\end{figure}

\begin{figure}
  \vspace{5mm}
	\begin{center} \includegraphics{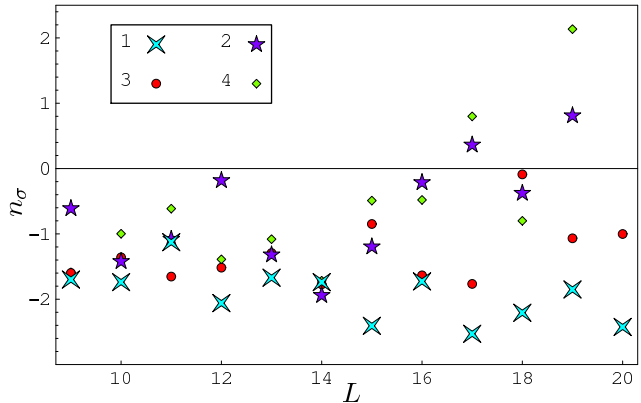} \end{center}
	\caption{We quantify here the number of standard deviations $n_\sigma$
	for which the value of the form factor of the KIC deviates from the RMT
	prediction.  This calculations were performed for 1, 2, 3 and 4 kicks.
	The prominent feature is that for one kick we are always near 2
	standard deviations away from the predicted result, systematically
	always undershooting the RMT result. For higher number of kicks the
	behavior is statistically as expected.} \label{fig:relativeerror}
\end{figure}

Perhaps a more clear picture is obtained after going into the time domain and
inspecting the form factor $K_2$ for a few kicks, which correspond to large
spectral ranges of $\Sigma^2$.  This regime is analogous to the
non-universality regime corresponding to the shortest classical periodic orbit
in quantum chaotic systems with well defined (semi)classical limit.  However,
we should not forget that our spin chain does not have any well defined
classical limit or semi-classical regime. Still, it seems that
$K_2(1/\tau_\H)$, $K_2(2/\tau_\H)$, etc., notably deviate from expectations of
COE of the same dimensions as the KIC propagator.  Indeed, in
fig.\ref{fig:K2at1} we show $K_2(1/\tau_\H)$ as a function of $L$, both for
individual quasi-momentum $k$ subspaces and the average over all relevant $k$,
and find very clear and systematic deviation from COE expectation $K_{2, {\rm
COE}}(1/\tau_\H) = 2/{\cal N}$. 
Furthermore, deviation of $K_2(1/\tau_\H)$ is much bigger that expected
COE fluctuation of $K_2(1\tau_\H)$ which can be computed as
$\sqrt{\<K_2(1/\tau_\H)^2\>_{\rm COE} - \<K_2(1/\tau_\H)\>^2_{\rm COE}} = 
2/{\cal N} + {\cal O}({\cal N}^{-2}).$
 Actually, in the limit ${\cal N}=\infty$, few
higher moments can be computed as well and COE distribution of $K_2(1/\tau_{\rm H})$
is conjectured to be exponential. 
In fig.~\ref{fig:relativeerror} we plot the relative deviation
of $K_2(t/\tau_H)$, for $t=1,2,3,4$, from expected COE average in terms of the
number of expected standard deviations.  It is clear that, at least for one
kick, the deviation is exceeding COE model significantly.  Namely we find the
deviation in the same direction for all different numbers of qubits $L$, and for
almost all $L$ it is exceeding two standard deviations.
We also find statistically significant deviations from COE for other values of
$t$, in particular for $t=3$, while the deviations for even arguments $t=2,4$
are less clear and conclusive.

A specialized reader may inquire for a comparison with the DODO random matrix
ensemble (which resembles semi-separable systems \cite{dodo}). We note that 
the deviations from the COE observed in this article cannot be accounted by the
semi-separable structure of the Floquet operator (\ref{eq:floquet}).

\section{Conclusions}

We have performed numerical calculations of large quasi-energy spectra of an
interacting multi-qubit system, namely the kicked Ising chain. No analytical
solution of the model is known, i.e.  the model is believed to be
non-integrable. Consistently with previous results in the literature
\cite{pineda:066120}, we find good agreement of  short-range level statistics
of the model with Dyson's ensemble of circular random matrices.  However, when
looking in detail at certain long-range spectral statistics, corresponding to
short-times, we find notable and significant deviations from random matrix
theory. This result reminds of non-universal regimes in semi-classical chaos
widely studied in the 1990's. However, this behaviour cannot be attributed to
periodic-orbits, since the system lacks any sensible definition of a
classical limit.   

We believe that the numerical results are intriguing and await for theoretical
explanation, perhaps in the direction of suggesting a new, abstract
semi-classical picture (perhaps along the lines of Ref.\cite{gibbons}).

\begin{acknowledgments}
  We acknowledge discussion with T. H. Seligman and F. Leyvraz.
  We are grateful for many insights gained in discussions with
  C. Bunge and E. Brady.
  The work of C.P. was supported by Direcci\'on General de
  Estudios de Posgrado (DGEP).  T.P. acknowledges
  support from Slovenian Research Agency (program P1-0044 and Grant
  No. J1-7347). CP thanks the University of Ljubljana and its
  group for Nonlinear Dynamics and Quantum Chaos
  for hospitality.  
\end{acknowledgments}

\bibliographystyle{apsrev}

\bibliography{paperdef,miblibliografia,specialbib}
\appendix

\section{Dimensions of the invariant subspaces}
\label{sec:dimhilbert}

Consider the computational basis $ S=\left\{ | m_0 m_1 \dots m_{L-1}\>,\, m_j
\in \{0,1,\dots,d-1\}\right\}$ of the Hilbert space of $L$ qudits $\mathcal
H=\mathcal H_\textrm{qudit}^{\otimes L}$. Let us generalize the translation
operator allowing the $m_j$'s to have integer values between 0 and $d-1$.  The
Hilbert space  $\mathcal H$ is foliated into $L$ subspaces $\mathcal H_k$ such
that for any $|\psi\> \in \mathcal H_k$, $T|\psi\>=\exp(2 \pi \imath
k/L)|\psi\>$ and $\mathcal H=\bigoplus_{k=1}^L\mathcal H_k$. Let $P_k$ be the
orthogonal projection operator, such that $P_k|\psi\> \in \mathcal H_k$, for any $|\psi\>
\in \mathcal H$.  An elegant solution to the problem of calculating $\dim
\mathcal H_k$ is presented here, following \cite{leyvrazprivate}.

We first study the condition under which a state $|n\> \in S$  is projected 
to the null ket (zero vector). 
Let $J$ be the smallest positive integer such that $T^J |n\>=|n\>$;
we call $J$ the primitive period of $|n\>$.
The action of the projection operator $P_k$ on $|n\>$ is
\begin{equation}
  P_k |n\> 
=\left(\sum_{j=0}^{\frac{L}{J}-1} (\varphi_{J,k})^j\right)
  \left(\sum_{j=0}^{J-1} \varphi_{j,k} T^j\right)|n\>,
\end{equation}
with $\varphi_{l,k}:=e^{-2\pi \imath \frac{l k}{L}}$. Notice that $P_k
|n\> = 0$ if and only if $\gamma=\sum_{j=0}^{\frac{L}{J}-1}
(\varphi_{J,k})^j=0$.  
Since $\gamma$ is the sum of a geometric series, its calculation is
straightforward: $\gamma=0$ if and only if $\varphi_{J,k} \ne 1$. As a conclusion we 
obtain that $P_k|n\> \ne 0$ if and only if $\frac{k J}{L} \in \mathbb{Z}$. 

Define the equivalence relation $\sim$ in $S$ as: $|n\>\sim|m\>$ if there
exists an integer $j$ such that $|n\>=T^j|m\>$.  Furthermore, if
$|n\>\sim|m\>$, then $P_k|n\> \propto P_k |m\>$, but if $|n\> \nsim |m\>$, then
$(\<n|P_k^\dagger) (P_k |m\>)=0$. In other words, elements in different
equivalence classes, are projected to orthogonal states.  Thus counting the
equivalence classes which are not projected to $0$ yields $\dim \mathcal
H_k$.

Let $\tilde N(J)$ be the number of equivalence classes whose elements have given primitive
period $J$. If we call $N(J)$ the number of
elements in $S$ that have primitive period $J$, then, $N(J)=J\tilde N(J)$.
Notice that 
\begin{equation}
  \label{eq:firstdiv}
  \sum_{\left\{J | \frac{L}{J} \in \mathbb{N}\right\}} N(J)=
  \sum_{\left\{J | \frac{L}{J} \in \mathbb{N}\right\}} J\tilde N(J)=d^L
\end{equation}
as the only allowed values for $J$ are the divisors of $L$.
Using M\"obius inversion formula we obtain
\begin{equation}
\tilde N(J)=\frac{1}{J} \sum_{\left\{m | \frac{J}{m} \in \mathbb{N}\right\}}
\mu\left(\frac{J}{m}\right)d^m.
\end{equation}
M\"obius function $\mu$ is defined over the positive integers as $\mu(1)=1$, $\mu(n)=0$
if $n$ is divisible by the square of a prime, and in any other case,
$\mu(n)=(-1)^p$ where $p$ is the number of prime factors of $n$. Then,
collecting our results,
\begin{equation}
  \label{eq:firstdim}
  \dim \mathcal H_k = 
  \sum_{\left\{ J | \frac{L}{J},\frac{k J}{L} \in \mathbb{N}\right\}} 
     \tilde N(J)
\end{equation}
since the only possible primitive periods $J$ are the divisors of $L$. 
However the value of $\dim \mathcal H_k$ is well approximated by $2^L/L$
for large values of $L$. 

\section{The optimal basis for diagonalization of KIC}
\label{basis}

To get the spectra used in this paper, it is crucial to develop an optimal
diagonalization scheme. Though the techniques relaying on Lanczos
method~\cite{effidiag} are fast they are not completely reliable.
They loose precision as soon as some eigenvalues are close enough. In our experience 
the Lanczos method
allows to obtain the full spectra for systems of up to 21 qubits, but the
intrinsic numerical error becomes comparable to the mean level spacing. Even 
for 18
qubits, the biggest numerical error in one level is already bigger than the
smallest inter-level spacing. As we are performing very precise tests we
require that our levels are highly reliable, making Lanczos a prohibitively
inexact method. We prefer using direct diagonalization with specialized
routines \cite{bunge}. 

Let
$U_\text{s,KI}=U_\text{kick}(\vec{b}/2)U_\text{Ising}(J)U_\text{kick}(\vec{b}/2)$
be the symmetrized version of $U_\text{KI}$, which however has the same spectrum
due to unitary equivalence. Using an appropriate basis is
important both to take advantage of the natural block diagonal decomposition of
$U_\text{s,KI}$ (due to symmetry $P$) and its symmetric character (due to
symmetry $\mathcal K$).  The basis is constructed as follows.  Let $|n\>$ be a
representative of a class defined by $\sim$ (see App.~\ref{sec:dimhilbert}), and
such that $P_k |n\> \ne 0$. If $\< n| R P_k |n\>=0$ then both $P_k |n \> \pm
\mathcal K R P_k |n \>$ are used as members of the basis.  If $\< n| R P_k |n\> \ne
0$ at least one of  $P_k |n \>  \pm \mathcal K R P_k |n \>$ is not the null ket,
and can be incorporated into the base. By choosing $|n\>$ from all different
classes we build a complete basis in $\mathcal H_k$.  Moreover, this basis is
invariant under $\mathcal K$ and is orthogonal. 

\end{document}